\documentclass[eqsecnum,aps,tightenlines]{revtex4}
\usepackage{epsfig,graphicx}

\begin{document}

\title{
Strangeness Saturation: Energy- and System-Size Dependence
}
\author{
{\sc J. Cleymans$^a$,
B. K\"ampfer$^b$,
P. Steinberg$^a$\footnote{Visiting Fulbright Professor on leave of absence from the Brookhaven National Laboratory, Upton, NY, USA},
S. Wheaton$^a$}
}
\address{
$^a$ Department of Physics, University of Cape Town,
Rondebosch 7701, Cape Town, South Africa\\
$^b$ Institut f\"ur Kern- und Hadronenphysik,
Forschungszentrum Rossendorf, PF 510119, D-01314 Dresden, Germany\\
}

\begin{abstract} 
Relativistic heavy-ion collisions lead to a final state which has a higher
 degree of strangeness saturation than those of elementary collisions. 
A systematic analysis of this phenomenon, based on the strangeness saturation 
factor, $\gamma_s$, is made for C+C, Si+Si and Pb+Pb collisions at 
the CERN SPS collider and for Au+Au collisions at RHIC energies. 
Strangeness saturation is shown to increase with the number of participants 
within a colliding system, at both CERN SPS and RHIC energies.
The saturation observed in central collisions of lighter nuclei  deviates 
from that seen in peripheral collisions of heavier nuclei with an 
equivalent participant number, which could be due to the difference in 
nuclear density.
\end{abstract}

\maketitle

\section{Introduction} 

It has been shown that statistical-thermal models are able 
to reproduce the multiplicities measured in 
relativistic heavy-ion collisions with remarkable success. This is accomplished with a very small number of parameters--
the temperature, baryon-chemical potential, $\mu_B$, and a factor measuring
the degree of strangeness saturation, $\gamma_s$.
As is now well known, there is very little difference between the temperatures 
observed in $p+p$ and relativistic heavy-ion collisions. The extracted 
strangeness saturation factor, $\gamma_s$, is however very different in 
$p+p$ and heavy-ion collisions. In this paper we focus on hadron 
multiplicities and extract the thermal parameters as a function of 
system-size and energy. 


The recent study in \cite{Bearden} (cf.\ table III therein)
impressively demonstrated that, with increasing system-size at SPS
energies, the strangeness saturation increases. In \cite{we} we 
have shown that at a beam energy of 158 AGeV, in 
collisions of lead-on-lead nuclei, the
strangeness saturation continuously increases with centrality.
However, strangeness (as measured by fully-integrated
kaon and antikaon multiplicities)
is clearly below saturation. A preliminary analysis 
\cite{Hirschegg,Budapest,Nantes} of the centrality
dependence at RHIC energy of $\sqrt{s}_{NN} = 130$ GeV points to 
a further increase of strangeness towards saturation for central
collisions of gold nuclei. 
An independent analysis \cite{NuXu,Kaneta} confirms this finding. 

This paper is divided into several sections. Firstly the system-size
 dependence of the thermal parameters is determined using 4$\pi$-yields 
from central C+C and Si+Si collisions~\cite{C_Si}, and centrality-binned Pb+Pb collisions~\cite{Sikler,Blume} 
at 158 AGeV at the CERN SPS. For comparison, centrality-binned 
mid-rapidity yields from Au+Au collisions at $\sqrt{s}_{NN} = 130$ 
GeV~\cite{PHENIX} 
and Pb+Pb collisions at SPS energy~\cite{Sikler} are analysed, despite 
the danger in applying the thermal model to yields in a limited 
rapidity window. 
Finally, the energy dependence of the thermal parameters is further 
elucidated by analysis of central Pb+Pb yields measured by NA49 at 
40, 80 
and 158 AGeV \cite{NA49_coll1,Mischke_QM02,NA49_coll2,NA49_ksi,Mischke}.  

\section{Analyses of Hadron Multiplicities} 

In the thermal model, hadron multiplicities can be described 
\cite{review,heavy_ions,PBM_qd,abundances_a,abundances_b} 
by the grand-canonical partition function
${\cal Z} (V, T, \vec \mu_i) = \mbox{Tr} \{
\mbox{e}^{- \frac{\hat H - \vec \mu_i \vec Q_i}{T}} \}$,
where $\hat H$ is the statistical operator of the system,
$T$ denotes the temperature, and $\mu_i$ and $Q_i$ represent the 
chemical potentials and corresponding conserved charges respectively.
In the analysis of $4\pi$-data, the net-zero strangeness and the baryon-to-electric charge ratio of the colliding nuclei constrain
the components of $\vec \mu_i = (\mu_B, \mu_S, \mu_Q)$.  
The particle numbers are given, in the Boltzmann approximation, by 
\begin{equation}
N_i^{\rm prim} = V g_i \gamma_s^{\left|S_i\right|}\int 
\frac{d^3 p}{(2\pi)^3} \, dm_i \,e^{-\frac{E_i - \vec \mu_i \vec Q_i}{T}}
\mbox{BW} (m_i),
\end{equation}
where we include phenomenologically a strangeness 
saturation factor, $\gamma_s$, with $\left|S_i\right|$ the number of valence 
strange quarks and anti-quarks in species $i$
\cite{Rafelski}
(i.e. $\gamma_s$ for the kaons and $\gamma_s^2$ for $\phi$) to account for 
incomplete equilibration in this sector, $E_i = \sqrt{\vec p^{\, 2} + m_i^2}$, and 
$\mbox{BW}$ is the Breit-Wigner distribution.
The particle numbers to be compared with experiment are
$N_i = N_i^{\rm prim} + \sum_j \mbox{Br}(j \to i) N_j^{\rm prim}$,
due to decays of unstable particles with branching ratios $\mbox{Br}(j \to i)$.  
For small participant numbers (typically $N_{\rm part}$ below 40),
one has to resort to a canonical or micro-canonical formalism
\cite{Redlich_Becattini,Keranen1}.

\subsection{System-size dependence}
\subsubsection{Analysis of fully-integrated yields}

\begin{figure}[tbh]
\centerline{
\includegraphics[width=12cm]{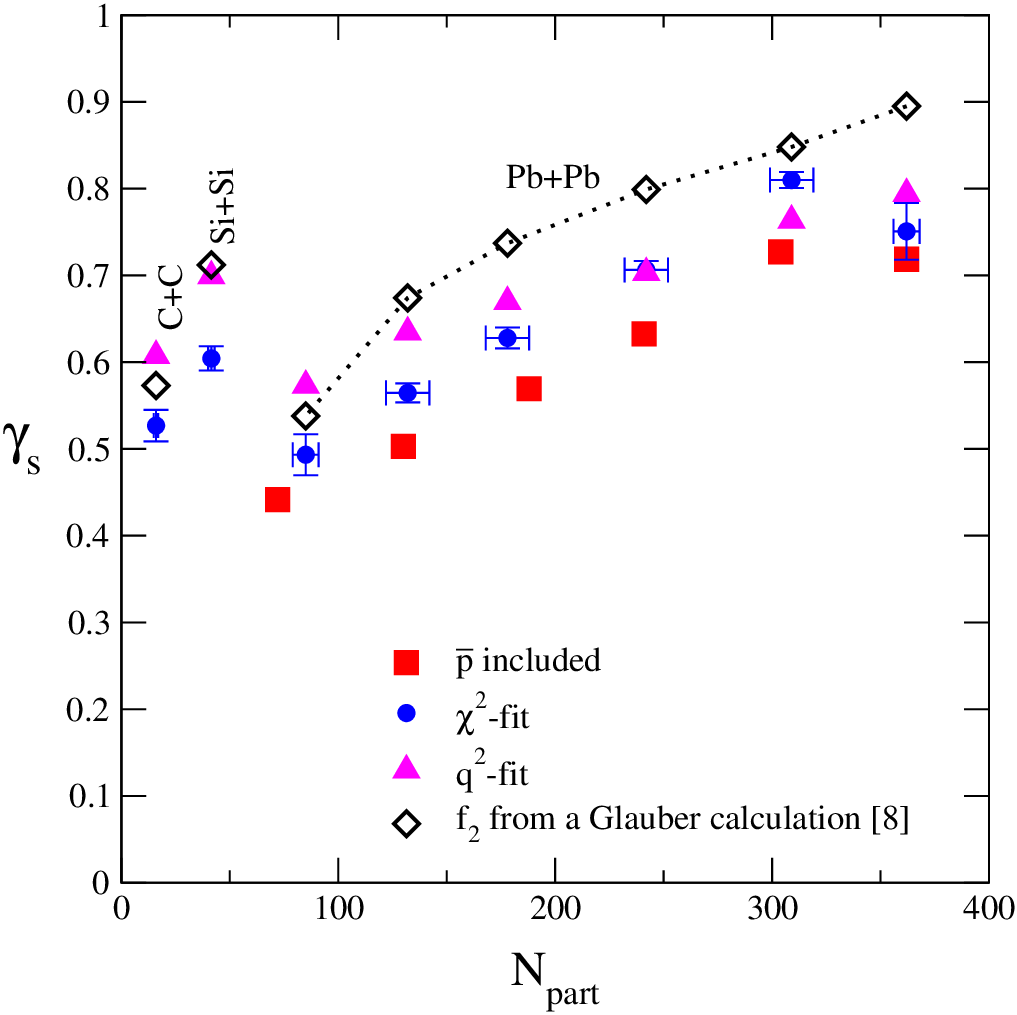}}
\caption{The system-size dependence of the strangeness saturation 
factor, $\gamma_s$, as extracted from centrality-binned Pb+Pb~\cite{
Sikler,Blume},
 and central C+C 
and Si+Si data~\cite{C_Si} under various fit conditions. The circles with 
error bars represent the results of our $\chi^2$-analysis assuming 
50\% feeding from weak decays, 
while the triangles show the results minimising the quadratic deviation (again 
assuming 50\% feeding). For comparison, the results of our earlier Pb system 
analysis \cite{Nantes}, with $\overline{p}$'s included in the fit, are 
included (squares). Also shown are the fraction of participants which 
underwent 
multiple collisions, $f_2$, as extracted from a Glauber 
calculation~\cite{C_Si} (diamonds).} 
\label{f_gammas_sys_size}
\end{figure}

\begin{figure}[tbh]
\centerline{
\includegraphics[width=12cm]{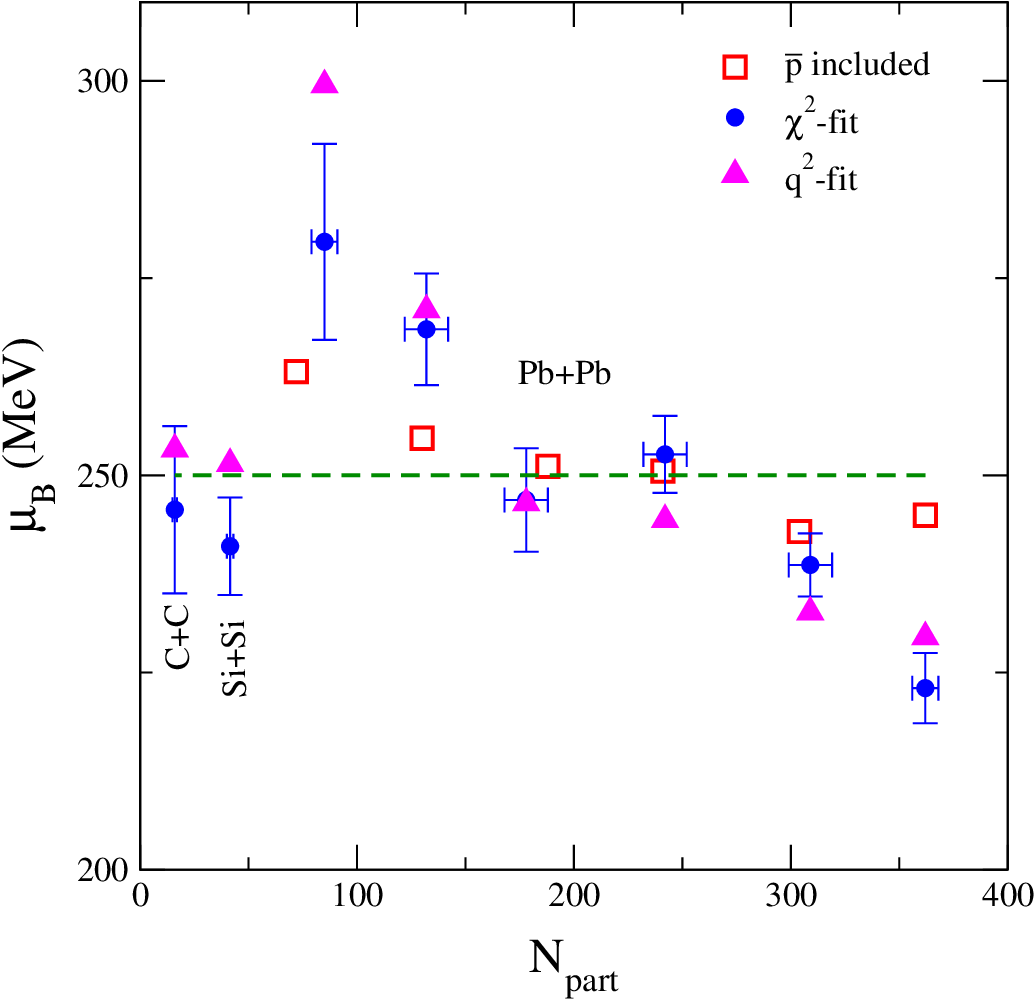}}
\caption{The system-size dependence of the baryon chemical potential, $\mu_B$, 
as extracted from centrality-binned Pb+Pb~\cite{Sikler,Blume}, and 
central C+C 
and Si+Si data~\cite{C_Si} under various fit conditions. The circles with 
error bars represent the results of our $\chi^2$-analysis assuming 
50\% feeding from weak decays, 
while the triangles show the results minimising the quadratic deviation (again 
assuming 50\% feeding). For comparison, the results of our earlier Pb system 
analysis \cite{Nantes}, with $\overline{p}$'s included in the fit, are 
included (squares). }
\label{f_muB_sys_size}
\end{figure}

In order to extract the system-size dependence of the thermal parameters we 
analyse $4\pi$-multiplicities of $\pi^\pm$, $K^\pm$, $\phi$ 
and $N_{\rm part}$ (taken as the sum over all baryons)
in 6 centrality bins in the reaction Pb+Pb~\cite{Sikler,Blume} 
(at 158 AGeV) and 
for central Si+Si and C+C collisions~\cite{C_Si} at the same energy.
 Our previous analyses \cite{we,Hirschegg,Budapest,Nantes} of the Pb+Pb system 
included $\overline{p}$ yields. They are excluded in this 
analysis in order that the Pb-, C- and Si systems be treated 
equivalently. No weak feed-down corrections have yet been applied to the
peripheral Pb-, or C- and Si systems \cite{Hoehne_private}. 
Due to the rather limited data set, the 
freeze-out temperature was fixed at 165 MeV, independent of centrality 
and colliding system. This is supported by a variety of fits to both heavy-ion and elementary collision systems in this energy regime~\cite{Bearden,we,heavy_ions,B1,B2}. Owing to the size of the C- , Si- and 
peripheral Pb systems, strangeness was treated canonically in all systems. As 
shown in~\cite{Keranen1}, for systems of this size at SPS energy, 
it is sufficient to treat the baryon- and charge content 
grand-canonically.  
The results are displayed in 
Figs.~\ref{f_gammas_sys_size} and \ref{f_muB_sys_size} with the specifics of 
each fit explained in the captions.

\begin{figure}[tbh]
\centerline{
\includegraphics[width=12cm]{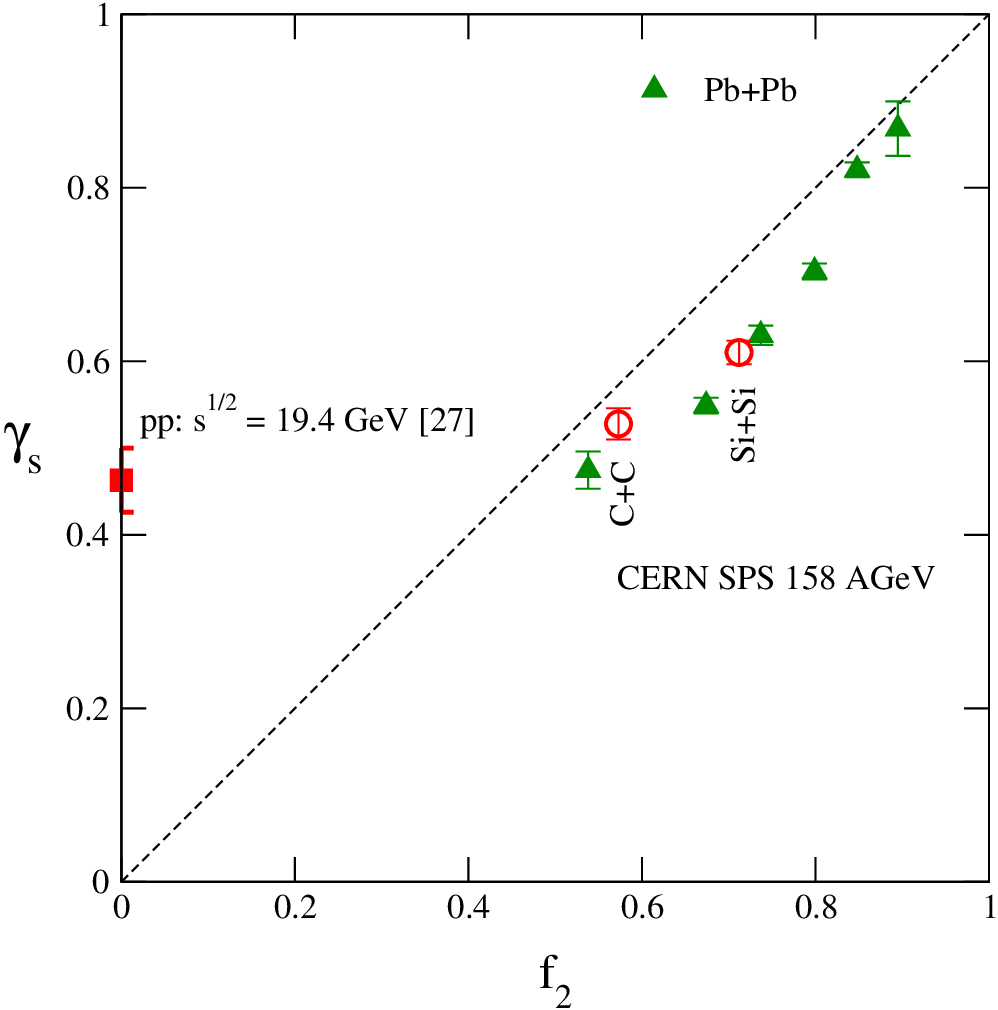}}
\caption{The strangeness saturation factor, $\gamma_s$, as extracted from 
centrality-binned Pb+Pb~\cite{Sikler,Blume} (triangles) and central C+C 
and Si+Si data~\cite{C_Si} (circles), as a function of $f_2$, the fraction of 
multiply-struck participants.  The results shown are those obtained with 
$\mu_B$ fixed at 250 MeV and $T$ at 165 MeV, assuming 50\% weak feed-down.
 For comparison, the strangeness saturation as extracted from 
p+p collisions at $\sqrt{s} = 19.4$ GeV~\cite{B2} is included 
(square). }
\label{f_gammas_f2}
\end{figure}

The strangeness saturation factor, $\gamma_s$, shows an increasing 
trend with collision centrality in the Pb+Pb system, except possibly over 
the two most 
central bins (see Fig.~\ref{f_gammas_sys_size}). It is also clear that 
the C+C and 
Si+Si systems lie above the trend suggested by the Pb+Pb points. 
This suggests that peripheral Pb+Pb collisions are not 
equivalent, with respect to strangeness saturation, to central 
collisions of lighter nuclei with the same 
participant number. 
In the C+C 
and Si+Si systems the baryon chemical potential is also lower than 
in the peripheral Pb+Pb bins (refer to Fig.~\ref{f_muB_sys_size}). It 
should be stressed that the only direct baryon information we have 
at our disposal in this analysis is the number of participants.
It would appear 
that, in the Pb+Pb system, $\mu_B$ decreases as the collisions 
become more central. 
As can be seen in 
Fig.~\ref{f_muB_sys_size}, the inclusion of $\overline{p}$ yields in 
the Pb analysis (squares) leads to a roughly centrality-independent baryon 
chemical potential of approximately 250 MeV. In order to 
investigate its interplay with $\gamma_s$, we fixed 
$\mu_B$ at 250~MeV in all 
collisions. This affected appreciably only the most central Pb+Pb 
bin, resulting in a monotonic increase in $\gamma_s$ within the Pb system. 
Thus, the drop in $\gamma_s$ over the most 
central bins of the Pb system is driven by $N_{\mathrm{part}}$ which, due to 
its small 
error relative to other species in the most central bin, is 
weighted heavily in a $\chi^2$-analysis. Since the errors, particularly in the C and Si systems have not yet been well-established~\cite{Hoehne_private}, we 
repeated the fits minimising the `quadratic 
deviation' \cite{PBM_qd} defined by:
\begin{equation}
q^2 = \sum_i{\left(M^{exp}_i-M^{model}_i\right)^2\over \left(M^{model}_i\right)^2}
\end{equation}
where $M^{exp}_i$ and $M^{model}_i$ are the experimental and 
model-predicted multiplicities of hadron species $i$ respectively.
The results of this analysis are 
shown in Figs.~\ref{f_gammas_sys_size} and \ref{f_muB_sys_size} as 
triangles.
Again $\gamma_s$ in the Pb+Pb system shows a monotonic trend, 
while the C+C and Si+Si 
systems still show a sizeable deviation from the peripheral Pb points. 
This cannot be 
attributed to rescattering, since it is more dominant in 
peripheral Pb+Pb than in central C+C and Si+Si reactions~\cite{Bass}. 
In~\cite{C_Si} it is shown that the strangeness enhancement, as 
measured by the ratios of strange to non-strange mesons, in these
systems scales with $f_2$, the fraction of participants which 
undergo multiple collisions. As shown in Fig.~\ref{f_gammas_f2}, 
$\gamma_s$ 
scales with this variable too. The strangeness saturation extracted from 
p+p collisions at $\sqrt{s} = 19.4$~GeV~\cite{B2} (denoted in 
Fig.~\ref{f_gammas_f2} by the square) suggests a strong flattening off 
of $\gamma_s$ for small systems. 
What is surprising is the 
approximate equivalence of $f_2$ and $\gamma_s$ for a number of points 
in Fig.~\ref{f_gammas_sys_size} ($f_2$ in this figure is denoted by the 
diamonds).\\  

\subsubsection{Mid-rapidity analysis}

When applying the thermal model to $4\pi$-data, many dynamical 
effects cancel out in ratios of the
 fully-integrated hadron yields \cite{cr}. 
In particular, 
effects due to flow disappear if the freeze-out surface
is characterized by a single temperature and chemical potential. 
When applying the thermal model to mid-rapidity data this is true 
only when the Bjorken model~\cite{bj} holds. Furthermore, in a limited 
rapidity window there 
is no guarantee that the total strangeness should be zero, nor 
that the baryon-to-charge ratio in this kinematic region should be 
set by the colliding nuclei. In a grand canonical approach this
affects 
the constraints on the chemical potentials. In the canonical formulation 
the `canonical suppression volume' (i.e. the volume in which quantum numbers 
are exactly conserved and that 
used to calculate the densities) and the `normalisation 
volume' (i.e. the volume required to convert densities to yields) 
are not necessarily the same.
A further complication arises in the treatment of decays. The 
experimental yields include feed-down from heavier resonances 
into stable, final-state particles. With mid-rapidity data, this 
requires careful consideration of the decay kinematics, as 
particles in a certain rapidity range will in general decay 
into particles in different kinematic windows. 

Despite these disclaimers 
we analyse the following mid-rapidity yields measured by the NA49 and 
PHENIX collaborations:  

i) NA49 mid-rapidity yields of $\pi^\pm$, $K^\pm$, $p$ and $\overline{p}$ in the reaction Pb(158 AGeV) + Pb in 6 centrality bins \cite{Sikler}.
 
 
ii) PHENIX mid-rapidity densities of $\pi^\pm$, $K^\pm$ and $p^\pm$
in the reaction Au + Au at $\sqrt{s} = 130$ AGeV in 5 centrality 
bins~\cite{PHENIX}. 

The PHENIX yields were not corrected for weak decays. PHENIX
estimate the probability for reconstructing protons from $\Lambda$ 
decays as prompt protons at 32\% at $p_T$ = 1 GeV/c \cite{PHENIX}. 
The PHENIX analysis was performed with both 0\% and 50\% feed-down 
from weak decays, while no weak feeding was included in the SPS 
analysis.
In all cases the grand-canonical formalism was applied, with the total 
strangeness set to zero but $\mu_Q$ fit as a free parameter. 

\begin{figure}[tbh]
\centerline{
\includegraphics[width=12cm]{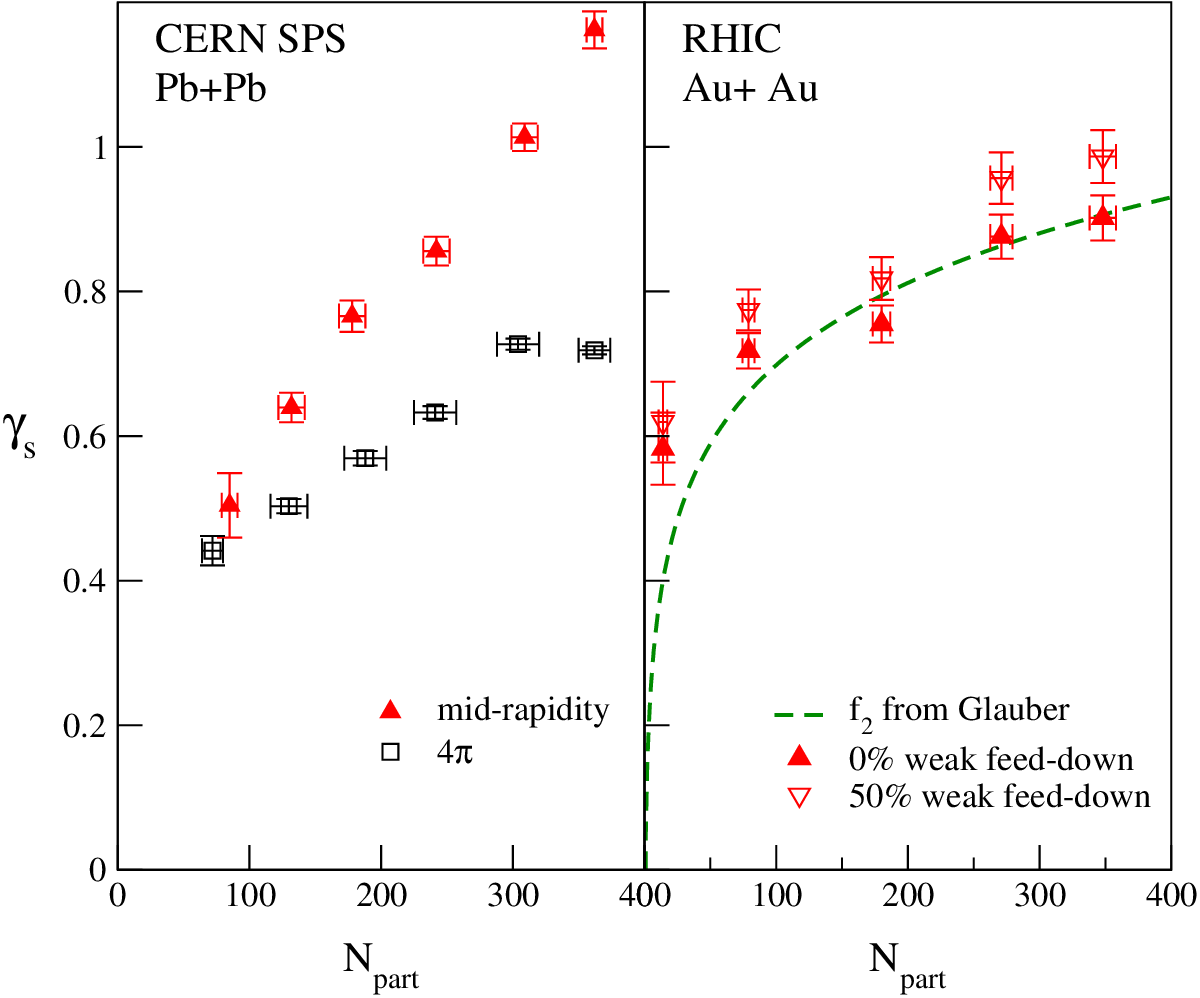}}
\caption{Left Panel: Comparison of the strangeness saturation factor, 
$\gamma_s$, extracted 
from mid-rapidity NA49 data~\cite{Sikler} (up triangles)  
with the results of our earlier analysis of NA49 4$\pi$-yields 
(squares)~\cite{Nantes}. 
Right Panel: The strangeness saturation observed in Au+Au collisions 
as extracted from PHENIX data~\cite{PHENIX}. The analysis was performed 
assuming 50\% weak feed-down (down triangles) and 0\% weak feed-down 
(up triangles). Also 
shown are the fraction of multiply-struck participants, $f_2$, obtained from our Glauber calculation (dashed line).}
\label{f_mid_rap}
\end{figure}

In Fig.~\ref{f_mid_rap} the system-size dependence of $\gamma_s$ at 
mid-rapidity is shown for SPS (left panel) and RHIC (right panel). 
In the SPS plot the results of our earlier analysis of fully-integrated 
NA49 yields~\cite{Nantes} are included for comparison. It is seen that $\gamma_s$, as extracted from the mid-rapidity NA49 data, 
is well above that obtained from the analysis of the fully-integrated 
NA49 yields. In order to exclude the possibility that the 
difference in the strangeness saturation extracted from the 4$\pi$- and 
mid-rapidity data analysed here is due solely to different 
strange hadrons included in the fits, the 4$\pi$ NA49 analysis was 
repeated with the hidden-strangeness $\phi$ excluded. In this way the 
two NA49 analyses are equivalent with respect to strange particles. 
This led to a slight decrease in 
$\gamma_s$ in the most peripheral bins, and thus an even larger 
difference between mid-rapidity and fully-integrated results. Thus, 
certainly at SPS energies, the degree of strangeness saturation is 
far higher in the central rapidity region. Included in the RHIC plot of 
Fig.~\ref{f_mid_rap} is $f_2$. This fraction of multiply-struck
participants parametrises the system-size dependence of the strangeness 
saturation factor in Au+Au collisions at RHIC energy remarkably well.\\          

\subsection{Energy dependence}

\begin{figure}[tbh]
\centerline{
\includegraphics[width=12cm]{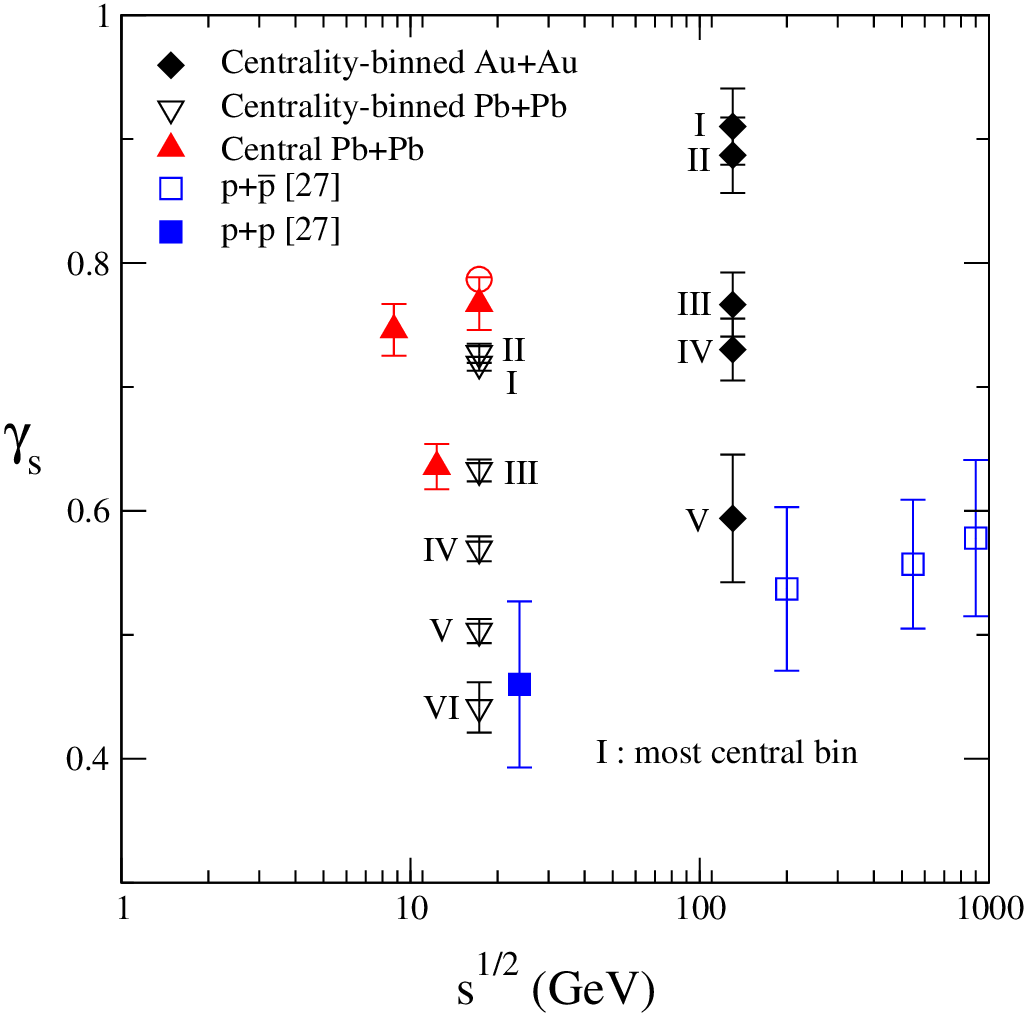}}
\caption{The energy dependence of the strangeness saturation factor, 
$\gamma_s$, extracted from central Pb+Pb collisions at 40, 80 and 
158 AGeV~\cite{NA49_coll1,Mischke_QM02,NA49_coll2,NA49_ksi} (up 
triangles), 
together with the results of our earlier analysis of centrality-binned 
Pb+Pb collisions at 158 AGeV (down triangles) and Au+Au collisions at 
RHIC (diamonds)~\cite{Nantes}. For comparison we show the results 
obtained from $pp$ collisions (filled squares) and $p\overline{p}$ 
collisions (open squares) at various 
energies~\cite{B2}.      
The open circle is extracted from mid-rapidity yields~\cite{
Mischke,NA49_ksi,NA49_coll1}.}
\label{f_gammas_e_dep}
\end{figure}

\begin{figure}[tbh]
\centerline{
\includegraphics[width=12cm]{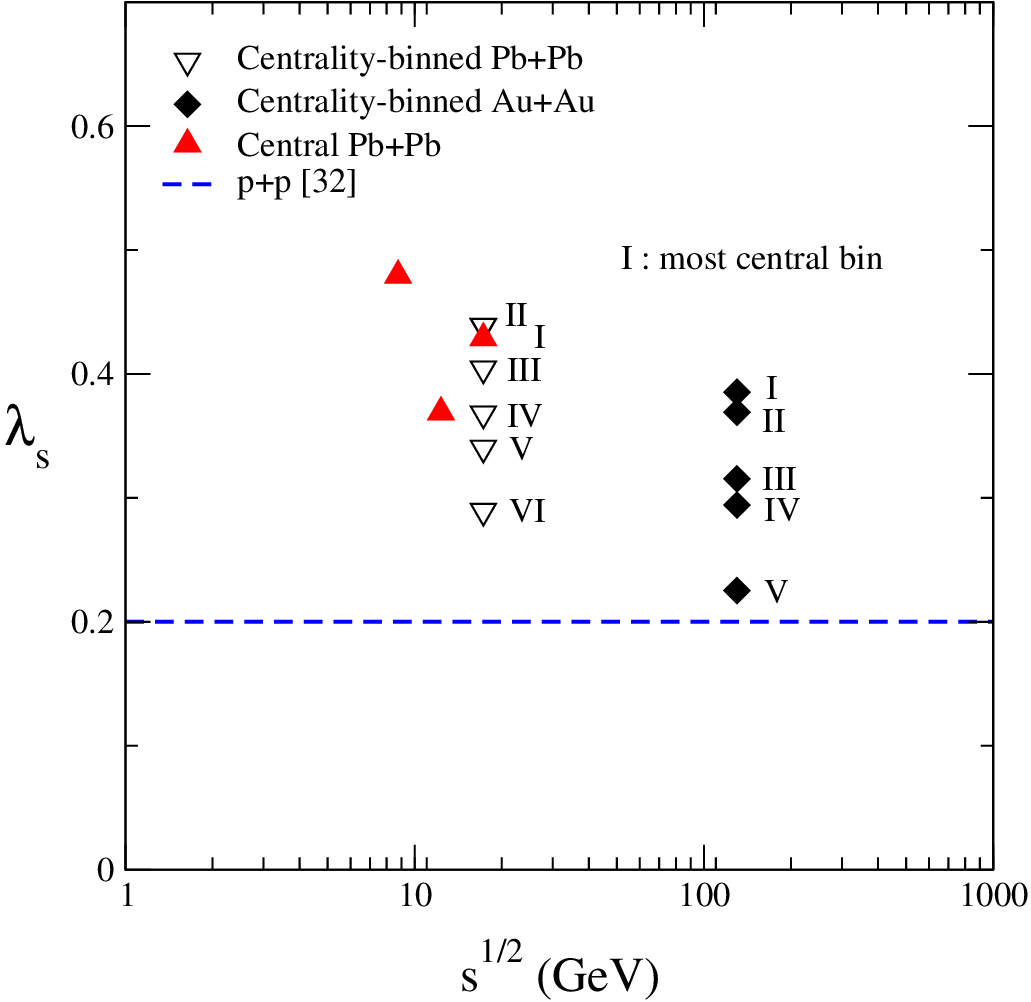}}
\caption{The energy dependence of the Wr\'{o}blewski factor, $\lambda_s$, 
extracted from central Pb+Pb collisions at 40, 80 and 
158~AGeV~\cite{NA49_coll1,Mischke_QM02,NA49_coll2,NA49_ksi} 
(up triangles), 
together 
with the results of our earlier analysis of centrality-binned Pb+Pb 
collisions at 158 AGeV (down triangles) and Au+Au collisions at 
RHIC (diamonds)~\cite{Nantes}. For reference we show the 
typical value 
of 0.2 extracted from $pp$ systems~\cite{B3}.}
\label{f_wrob_e_dep}
\end{figure}

In order to further investigate the energy dependence of the thermal 
parameters we analyse the 
fully-integrated yields of $\pi^\pm$, $K^\pm$, $\Lambda$ and 
$\overline{\Lambda}$ in central Pb+Pb collisions at 40, 80 and 158 AGeV, 
supplemented with $K_s^0$, $\Xi^-$, $\overline{\Xi}^+$ and 
$\phi$ multiplicities at 158 AGeV \cite{NA49_coll1,Mischke_QM02,NA49_coll2,NA49_ksi}. 
In Figs.~\ref{f_gammas_e_dep} and \ref{f_wrob_e_dep}, $\gamma_s$ and 
the Wr\'{o}blewski factor~\cite{Wr}, $\lambda_s$, which measures the ratio of newly created $s\bar{s}$ pairs to newly created non-strange valence quark pairs at the primary hadron level:
\begin{equation}
{\displaystyle \lambda_s = {2\langle s\bar{s}\rangle \over \langle u\bar{u}\rangle + \langle d\bar{d}\rangle}},
\end{equation}
 are displayed as a function 
of the collision energy. Included in the figures are the results of our earlier system-size analysis of the Pb+Pb system at CERN SPS and the Au+Au system at RHIC~\cite{Nantes}. It should be 
noted that the centrality cuts on the most central Pb+Pb collisions are 
slightly 
different at the various 
energies (7.2\% at 40 and 80 AGeV, and 5\% at 158 AGeV). In view of
the system-size dependence extracted in the previous section, this 
will raise the 40 and 80 AGeV points relative to the 158 AGeV point.   
In Fig.~\ref{f_wrob_e_dep} one observes that $\lambda_s$ for central 
collisions decreases with 
collision energy from 40 AGeV (in agreement with \cite{AGS_peak}). 
Within a given collision system, it shows a systematic increase with 
participant number, while remaining above the typical value of 0.2 
seen in $pp$ collisions~\cite{B3}. With respect to $\gamma_s$ there 
is fairly good agreement between the 
most peripheral heavy-ion bins and the results from elementary systems 
of comparable energy~\cite{B2}.         

For comparison we also show the value of $\gamma_s$ 
extracted from 
mid-rapidity Pb+Pb data~\cite{Mischke,NA49_ksi,NA49_coll1} at 158 AGeV 
(open circle) in Fig.~\ref{f_gammas_e_dep}. As can be seen, 
the strangeness saturation in the mid-rapidity region is clearly 
greater than that averaged over 4$\pi$. 

\section{Summary} 
In conclusion, the strangeness saturation factor, $\gamma_s$, has been 
shown to increase with participant number in the Pb+Pb system at the 
CERN SPS as well as the Au+Au system at RHIC. Central collisions of C+C 
and Si+Si at SPS energies deviate, with respect to strangeness 
saturation, from peripheral Pb+Pb collisions. However, $\gamma_s$ is 
seen to scale with the fraction of multiply-struck participants, $f_2$. 
In fact, $f_2$ remarkably tracks the $N_{\rm part}$-dependence of 
$\gamma_s$ as extracted from mid-rapidity yields in Au+Au collisions at RHIC. 
Where both mid-rapidity and 
fully-integrated data was available, the degree of strangeness 
saturation observed at mid-rapidity was found to be consistently higher than that extracted from 4$\pi$-data.\\

\section*{Acknowledgments} We acknowledge  useful correspondence with C. H\"ohne
and M. Ga\'zdzicki. One of us (S.W.) acknowledges the financial assistance of 
the National Research Foundation (NRF) of South Africa.

\newpage
\newpage
\newpage
\newpage
\newpage
\newpage
\end{document}